\documentclass[twocolumn, notitlepage,aps,showpacs,floats,amssymb,amsmath,floatfix,groupedaddress,superscriptaddress,aps,pre]{revtex4-2}
\usepackage{amsfonts,amssymb,stmaryrd,latexsym,amsmath,braket}
\usepackage{graphicx} 
\usepackage{booktabs}
\usepackage{comment}
\usepackage{newtxtext, newtxmath} 
\usepackage{slashed}
\usepackage{bm}
\usepackage{appendix}
\usepackage{enumitem}
\usepackage{multirow}
\usepackage{array}
\usepackage{tabularx}
\usepackage{placeins}
\usepackage[dvipsnames]{xcolor}
\usepackage{mhchem}

\usepackage[colorlinks=true,linktocpage=true,citecolor=blue,urlcolor=blue,linkcolor=blue]{hyperref}
\begin{document}

\title{Charge Regulation Mediated Interaction of Amphoteric Nanoparticle Surfaces}
\author{Saurav Tyagi}
\affiliation{Department of Physics, Indian Institute of Technology, Jodhpur 342030, Rajasthan, India}

\author{Kamal Tripathi}
\affiliation{Université Grenoble Alpes, CNRS, Grenoble INP, 3SR, 38000 Grenoble, France}

\author{Saikat Chakraborty}\email{saikat.chakraborty@vit.ac.in}
\affiliation{Centre for Functional Materials, Vellore Institute of Technology, Vellore, India}


\author{Sunita Kumari} \email{sunita@iitj.ac.in}
\affiliation{Department of Physics, Indian Institute of Technology, Jodhpur 342030, Rajasthan, India}

\date{\today}

\begin{abstract}
Charge regulation refers to the ability of ionizable biomolecules and their counterparts to adjust their ionization state in response to external perturbations, including changes in pH, ionic strength, or electrostatic interactions with neighboring charged species. Here, we use explicit-ion molecular dynamics / charge-regulation Monte Carlo simulations to compute the interaction force between two spherical, charge-regulated nanoparticles as a function of separation, and compare symmetric (chemically identical) and asymmetric (chemically complementary) surface configurations against the constant charge approximation. We find that charge regulation produces up to an order-of-magnitude stronger short-range attraction than constant charge for asymmetric nanoparticle pairs, particularly at low electrolyte concentration, and that charge regulation qualitatively alters the interaction between symmetric, chemically identical nanoparticles, driving a crossover from net repulsion to net weak-attraction as ionic strength decreases. This transition is not observed under the constant charge assumption which shows that charge regulation is not a minor correction but can change both the strength and the sign of nanoscale electrostatic interactions, with implications for the assembly and stability of charge-regulated colloidal systems.
\end{abstract}

\maketitle


\section{Introduction}
Charge fluctuations have always been of importance in our life due to the prevalence of charges in most natural phenomena\cite{muthu23qmacromols, wang24qdrivenassembly,Wang2026, levin02electrostatic,levin02electrostatic}. These charge fluctuations are not mere perturbations around some fixed equilibrium state; they are intrinsic to the system and deeply coupled to its structural and thermodynamic behavior\cite{lund13biomol, podgornik18gencr}.  Several biomolecules such as proteins, DNA, and even engineered charged moieties tune their ionization state in response to external physicochemical signals. This phenomenon is generally referred to as charge regulation (CR)\cite{podgornik18gencr,muthu23qmacromols, yuan22crzwit, curk22acsimcr, Yuan2024-ql, kumari24screencrmion, kumari24screencrmion, yadav26crpe, Yuan2024-ql, ruixuan23mioncr, lund13biomol, veider24npamr}. In this context, tiny sized nanopartilces (NPs)  are particularly interesting due to their large surface-to-volume ratio and tunable surface functionalities, which significantly enhance their bio-reactivity\cite{dela12elisa, lin20phcynp}. This sensitivity to their surrounding milieu is exploited in a variety of applications ranging from pH-triggered drug release\cite{gao10npdd, otsuka03pegnp, han20zwitdd}, to the nanofluidic devices\cite{jiang11crnpore, ritt22crnanochannels} and light-responsive nanoemulsions\cite{glikman24crlight}. CR is often linked to complex structural and functional behaviors. For instance, CR leads to a loosely packed assembly of charged NPs, contrary to the constant charge (CC) approximation\cite{yuan22crzwit}. It may even provide the driving force behind NP crystallization \cite{bian21npassembly}. Therefore, a deeper understanding of CR mechanisms is essential for optimizing the potential of charged NPs in diverse applications, and for a fundamental understanding of their interactions\cite{lund13biomol,curk21crnpasmb}.

The historical study of electrostatic interactions began with planar charged surfaces, where the Poisson-Boltzmann (PB) theory provided a mean-field description of the electric double layer formed by mobile ions around charged interfaces\cite{vasileva23electrolytepore, blossey23pbeq, podgornik18gencr, markovich16crgen}. This theory, with its CC and constant potential (CP) boundary conditions, laid the foundations for the celebrated Derjaguin–Landau–Verwey–Overbeek (DLVO) theory, which quantitatively explained the stability and aggregation of colloidal dispersions\cite{french10longnano, verwey1947lyophobic, trefalt17asymcoll}. However, DLVO theory does not take into account electrostatic ion-ion and ion-surface site correlations\cite{verwey1947lyophobic}. Moreover, traditional PB theory breakdowns while dealing with complex geometries, high surface charge densities, or high ion valencies\cite{levin02electrostatic,colla24crnpmulti,Kumari2022}.

The concept of CR was first introduced by Linderstr{\o}m-Lang in the context of protein ionization \cite{linderstrom1924ionization}, and later formalized by Kirkwood and Shumaker for polyelectrolytes\cite{kirkwood1952protein}. In the 1970s, Ninham and Parsegian\cite{ninham1971ionizable} proposed a nonlinear PB equation and excluded the individual nature of ions and functional groups, which has been shown to play a crucial role in the CR mechanism\cite{bakhshandeh19crcolloidtheory, bakhshandeh20crcolloidsolution, bakhshandeh20crmetalnp, gomez21latticecr}. Subsequent theoretical developments employed the law of mass action\cite{chan1975potcrampho}, partition function and free energy approaches\cite{markovich14freenrgcr,markovich16crgen}, and the Frumkin–Fowler–Guggenheim isotherm to account for nearest-neighbor interactions and multiple adsorption equilibria\cite{ruixuan23mioncr, kumari24screencrmion,kandari26screencrzwit}. For metal NPs, the CR boundary condition is often approximated by the CC condition, but significant deviations arise for systems with large charge asymmetry or strong environmental responsiveness\cite{bakhshandeh20crmetalnp, glikman24crlight}. 

While there is growing recognition of CR effects on two-body interactions\cite{chan1976edlampho, mccormack1995edlfreenrg, behrens1999electrostaticcolloid, biesheuvel04freenrgiondl, borkovec08edlxcr, chan06edlcr, boon11crscreenpatch, adzic14fieldtheorycr, adzic15crthermfluc, adzic16mionmulti, kubincova20interfacialsolvation, obstbaum22thermocr}, there is a significant lack of quantitative pair-wise force calculations, particularly for spherical aggregates at the nanoscale and microscale under CR conditions\cite{krishnan17qmionpe, behjatian22likecharge, ruixuan23mioncr}. This is especially true for pairs of identical (symmetric) spherical aggegates versus two complementary, oppositely-functionalized (asymmetric) surfaces, where CR is expected to couple most strongly to the sign and range of the resulting interaction. This gap hinders a complete understanding and accurate prediction of how these particles behave and interact in various applications.

In this work, we address this gap by performing a molecular dynamics simulations aided with charge regulating Monte Carlo steps (MD/CR-MC) with explicit ion addition/deletion\cite{curk22acsimcr}. We compute the total interaction force between two spherical, charge-regulated NPs as a function of their center-to-center separation. We also quantify surface ion adsorption. We consider both symmetric and asymmetric surface charge distributions, and compare the resulting forces and ion-adsorption behavior against the CC approximation across ion-rich and ion-poor electrolyte conditions. We find that CR produces substantially stronger short-range attraction than CC in the asymmetric case. The enhancement reaches up to an order of magnitude at low solution concentration. Moreover, CR can qualitatively change the nature of the interaction in the symmetric case, driving a crossover from net repulsion to net weak attraction as concentration decreases, a transition absent under the CC treatment. Our results demonstrate that CR is not simply a quantitative correction but can qualitatively alter the sign and magnitude of NP-NP interactions, underscoring the need to move beyond the CC approximation when modeling charge-regulated nanoscale assemblies.

This paper is organized as follows. Section~\ref{sec:simulation} outlines the numerical models and computational methods used. In Section~\ref{sec:results}, we describe the results of the simulation study. Finally, we present conclusions in Section~\ref{sec:conclusions}. 

\section{MODEL AND SIMULATION}
\label{sec:simulation}
\begin{figure}
\centering
\includegraphics[width=0.49\textwidth]{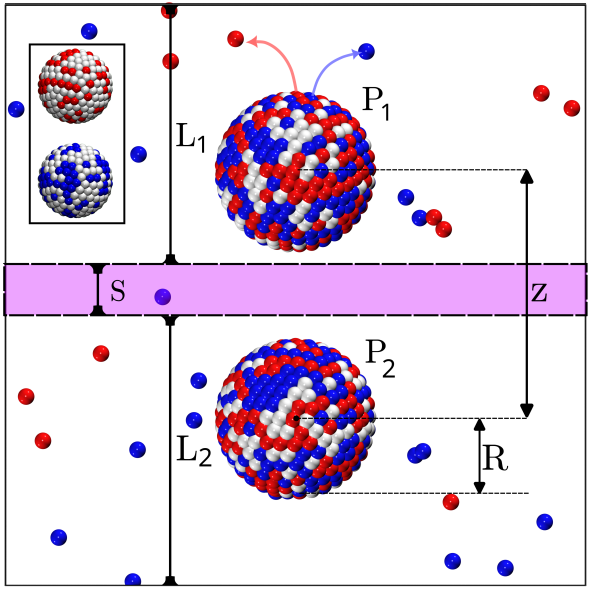}
\caption{\textbf{Schematic of simulation setup:} A cubic simulation box contains two rigid spherical NPs, $\text{P}_1$ \& $\text{P}_2$, each of radii $\text{R}$ at center-to-center distance $z$, surrounded by explicit monovalent ions. The box is divided into three imaginary regions for force calculations: the upper region $\text{L}_1$, the lower region $\text{L}_2$, and the highlighted mid-slice $\text{S}$ of thickness $1.5\sigma$. The curved arrows in faded red/blue depict the association/dissociation reactions that occur as part of the charge regulation scheme. All quantities are reported in reduced units. While the main schematic displays a symmetric distribution of charged species, the inset image shows an asymmetric distribution. 
}
\label{fig1}
\end{figure}

As illustrated in Fig.~\ref{fig1}, we consider a system of two charge-regulated spherical nanoparticles (NPs), $P_1$ and $P_2$, placed at a center-to-center separation $z$ and immersed in the solution of an explicit monovalent electrolyte. The electrolyte ions surrounding the NPs are categorized by charge into two species: positive particles ($X^+$ and $H^+$) shown in blue, and negative particles ($X^-$ and $OH^-$) shown in red. The simulations are performed in a cubic box of size $200\sigma$ under periodic boundary conditions. Here we set the distance unit $\sigma$ to be equal to the Bjerrum length $\ell_B= q^2/4\pi \varepsilon\varepsilon_0k_B T$. It is a distance at which the electrostatic interaction between two units of charges ($q$) equals the thermal energy ($k_B T$). In our simulation, we kept $\ell_B = 0.72$ nm for water at room temperature. The terms $\varepsilon$ and $\varepsilon_0$ are the dielectric constant of the solvent and the vacuum permittivity, respectively. Each NP carries a set of 256 titrable surface sites. Known solutions to the Thomson problem\cite{thomson1904tp, wales06tpsol}, which concerns the uniform distribution of points on a sphere, are used to place the 256 titrable sites uniformly on the surface. The diameter of these surface sites and mobile ion is set at $\sigma$. These groups are attached rigidly to each of the spherical cores, resulting in an effective outer NP radius of $R=4\sigma$. These surface sites are either acidic or basic with acid and base dissociation constants $pK_{a}$ and $pK_{b}$, respectively. Here, an acid refers to surface sites that can acquire a charge of $q=0$ or $q=-1$, while a base refers to surface sites that can acquire a charge of $q=0$ or $q=+1$. The charged groups on the surface can adjust their ionization state to interact with ions in the solution, all while maintaining an overall charge balance within the system. Acidic and basic sites follow the equilibrium reaction $\ce{A <=>[pK_a] A^- + X^+}$ and $\ce{B <=>[pK_b] B^+ + X^-}$ respectively, where $X^+$ and $X^-$ are generic free ions. Free ions in the electrolyte, with ion activity constant $pK_{\phi}$, follow the equilibrium $ \ce{ $\phi$ <=>[pK_{\phi}] X^{+} + X^{-}}$\cite{curk21crnpasmb, curk22acsimcr}.


We follow the hybrid MD/CR-MC scheme to simulate our system\cite{curk22acsimcr}. We use the velocity-Verlet algorithm to update the positions of the particles\cite{curk21crnpasmb, curk22acsimcr}.  In addition, every $N_{MD}$ MD steps, $N_{MC}$ MC moves are employed, that change the charge states of the ionizable particles on the surfaces. Corresponding free ions are inserted or removed to maintain the overall charge neutrality of the system. The free ions are coupled to an implicit reservoir through effective ion activities, represented by $pI_p$ and $pI_m$ for positive and negative ion species, respectively. Both are defined in the  $-\log 10$ representation. For example, $10^{-{pI_p}} = 10^{-{pH}} + 10^{-{pI_{s^+}}}$, where $pI_{s^+}$ denotes the concentration of added salt cations\cite{curk22acsimcr}. We perform all calculations in a $pH = 7.0$ environment. The free ion activities $pI_p=pI_m\approx-log(c)$, where $c$ is the reduced molarity of the free ions in the box, are set at $-log(c) \approx 3$ or $c\approx10^{-3}$ for an ion-rich environment and $-log(c) \approx 6$ or $c\approx10^{-6}$ for an ion-poor environment. Accordingly, we label them as higher and lower concentration runs. 

To compare with the CC case, the average ionization states observed in the CR simulation are calculated, and these average charges $\langle Q\rangle$ for each NP are then assigned as fixed charges to the NP pair in a separate CC simulation, so that each NP has a constant charge of $\langle Q\rangle$. Both conditions are simulated using MD, employing the same numerical integration scheme and system parameters. All simulations were performed with a time step of $0.005\tau$, using a Langevin thermostat with a damping constant of $1.0\tau$ to maintain the temperature at $T=1.0$. 

The excluded volume interactions are calculated using the expanded Lennard-Jones (LJ) potential, which ensures that the finite size and impenetrability of particles are accurately represented in the simulation and given by, 
\begin{equation}
    U_{LJ}(r_{ij})=
    \begin{cases}
      4\varepsilon \left[ \left(\frac{\sigma}{r_{ij}-\Delta}\right) ^{12}  - \left(\frac{\sigma}{r_{ij}-\Delta}\right) ^{6} \right] , & r_{ij} \leq r_c^{LJ}~,\\
      0~, & r_{ij}>r_c^{LJ}~,
    \end{cases}
    \label{lj}
\end{equation}
where $\epsilon$ is the depth of the potential well, $r_{ij}$ is the center-to-center separation between particles $i$ and $j$. $\Delta$ is the expanded distance and the cut off $r_c^{LJ} = \Delta + 2^{1/6}$. The expanded LJ potential parameters for different particle types are quoted in Table~\ref{tab:table1}.
\begin{table}[ht]
\caption{\label{tab:table1} LJ interaction parameters between different types of particles; these are categorized into small particles with radius $\sigma/2$ (free charge carriers and surface sites) and NPs with radius $R$. }

\begin{tabularx}{0.3\textwidth} { 
  | >{\centering\arraybackslash}X 
  | >{\centering\arraybackslash}X | }
 \hline
Particle Prototype & Value of $\Delta$ \\
 \hline
$R \leftrightarrow R$ & $2R - \sigma$ \\
\hline
$R \leftrightarrow \sigma/2$ & $R - \sigma/2$ \\
\hline
$\sigma/2 \leftrightarrow \sigma/2$ & $0$ \\
\hline
\end{tabularx}
\end{table}

The Columbic interaction between charges is computed by,
 \begin{equation}
U_{\mathrm{Coul}}(r_{ij})=
\begin{cases}
  \frac{q_i q_j}{4\pi \epsilon_0 \epsilon r_{ij}}, & r_{ij} \leq r_c^{\mathrm{Coul}}~,\\
      0~, & r_{ij}>r_c^{\mathrm{Coul}}~,
      \label{Coul}
\end{cases}
\end{equation}

 $q_i$, $q_j$ are the charges of the interacting particles and $r_c^{Coul}$ is the cut off distance for the long range Coulombic interaction. 


\subsection*{Total force between two Nanoparticles}

To analyze the total force acting between two spherical NPs $P_1$ and $P_2$, we consider a fictitious plane $``S"$ of size $ 200\sigma\times 200\sigma\times \sigma$, which is conveniently placed at the midpoint between $P_1$ and $P_2$, see Fig~\ref{fig1}. The sub-volumes $L_1$ and $L_2$ extend over the upper and the lower halves of the system so that the mid-slice $S$ coincides with the mid-plane. As discussed in detail in Refs~(\cite{podgornik1995colloidpe, yadav26crpe}), the force ($F_{z}$ ) acting along the radius vector joining the $P_1$ and $P_2$ is obtained by,
\begin{eqnarray} \nonumber
F_{z} &=& k_BT \int_{S} d^2 \mathbf{r}\rho_1(\mathbf{r}) \\
&+& \int_{L_1} \int_{L_2} d^3\mathbf{r_1} d^3\mathbf{r_2}\rho_2(\mathbf{r_1},\mathbf{r_2}) \times f_z(\mathbf{r_1},\mathbf{r_2})
\label{total}
\end{eqnarray}
 The first term of Eq.~\ref{total} resembles ideal osmotic pressure and represents the exchange of momentum between $L_1$ and $L_2$ due to the movement of ions across the $S$. The second term $f_z(\mathbf{r_1 ,r_2} )$ accounts for the microscopic force acting between a particle located at $r_1 \in L_1$ and at $r_2 \in L_2$ in the direction of the normal to the dividing surface $S$. $\rho_{1}(\mathbf{r})$ refers to the particle density of mid-slice and $\rho_2(\mathbf{r_1},\mathbf{r_2})$ is the particle density of the rest system. The microscopic force $ f_z(\mathbf{r_1 ,r_2} )$ primarily consist of two components. The first component arises directly from the Coulomb contribution ($F_{dir}$) between two CR NPs. The second part stems from the correlational contribution ($F_{corr}$). Therefore following the framework established via plane-parallel geometry\cite{podgornik1995colloidpe,granfeldt1991mccolloidpe, yadav26crpe} the total force are then assumed to have the following form:
\begin{eqnarray}
F_{z}= F_{osm}+ F_{dir}+F_{corr}
\label{eq:f_tot}
\end{eqnarray}

The osmotic force is calculated from the number of mobile ions within the slice $S$ according to
\begin{eqnarray}
    F_{osm}
    =
    k_{\mathrm{B}}T
    \left(
        n_S-n_b
    \right),
    \label{eq:osm}
\end{eqnarray}
where $n_S$ is the number of mobile ions within the mid-slice $S$ and $n_b$ is the corresponding baseline value in the absence of the NPs. 


The direct interaction force, $F_{dir}$, accounts for the explicit interactions between $P_1$ and $P_2$. It is obtained by summing the Coulomb force ($ F_{dir}^{\mathrm{coul}}$) and expanded LJ interaction ($F_{dir}^{\mathrm{LJ}}$) and given by, 
\begin{eqnarray}
    F_{dir}
    =
    F_{dir}^{\mathrm{coul}}
    +
    F_{dir}^{\mathrm{LJ}}.
    \label{eq:dirtot}
\end{eqnarray}


This correlational force, $F_{corr}$, arises from interactions between the mobile ions and the NP, as well as from the interactions between mobile ions located on opposite sides of the $S$, and is evaluated as follows
\begin{eqnarray}
    F_{corr}  = F_{P_1\rightarrow L_2} + F_{L_1\rightarrow P_2}  + F_{L_1\rightarrow L_2},
    \label{eq:f_corr}
\end{eqnarray}
where $F_{P_1\rightarrow L_2}$ is the force on $P_1$ due to the mobile ions in $L_2$, $F_{L_1\rightarrow P_2}$ is the force on the mobile ions in $L_1$ exerted by $P_2$, and $F_{L_1\rightarrow L_2}$ is the force between the free ions in $L_1$ and $L_2$.
We have calculated all force components (Eq.~\ref{eq:f_tot}) directly from the MD simulations using the compute group/group command in LAMMPS\cite{LAMMPS}. This includes both real-space and reciprocal-space electrostatic forces, which are processed via the PPPM solver with a force accuracy of $10^{-3}$.

 

\section{Results and Discussion}
\label{sec:results}


\subsection{Asymmetric Nanoparticles}

\begin{figure*}
    \centering
    \includegraphics[width=0.6\textwidth]{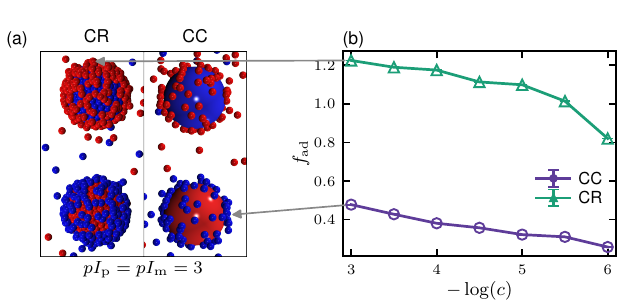}
    \caption{\textbf{CR compared with CC Baseline for Asymmetric NPs}
    (a) Snapshots of the system at $pI_p=pI_m=3 (c\approx 10^{-3}$)  for a fixed  NPs separation of  $z=15\sigma$: CR (left) and CC (right)
    (b) Variation of adsorption fraction ($f_{ad}$) with electrolytic concentration $-\log(c)$ at separation $15\sigma$, at $pH=7.0$ with acid and base dissociation constants $pK_a=pK_b=3$.}
    \label{fig2}
\end{figure*}

\begin{figure*}
    \centering
    \includegraphics[width=\textwidth]{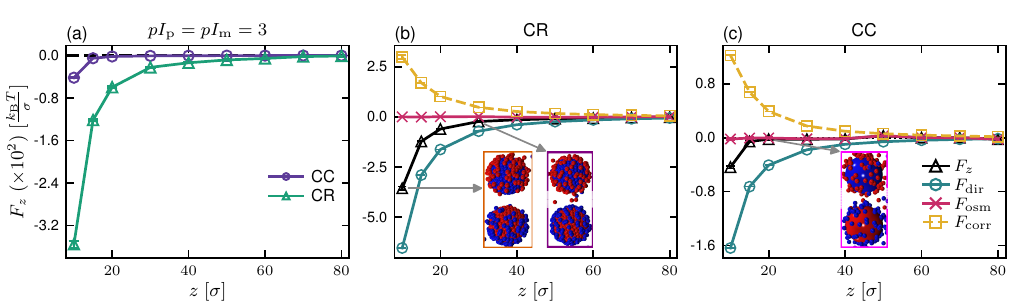}
    \caption{\textbf{Force decomposition for Asymmetric NPs}
    (a) Comparison of effective interaction force ${F}_{z}$ as a function of separation $z$ for CR and CC, and component wise breakdown of forces for (b) CR, insets are a snapshot of CR system at $z=10\sigma$, $z=30\sigma$, and (c) CC, inset is a snapshot of CC system at $z=20\sigma$, all at $pI_p=pI_m=3$ in the electrolyte at acid and base dissociation constants $pK_a=pK_b=3$ for the surface sites.}
    \label{fig3}
\end{figure*}

\begin{figure*}
    \centering
    \includegraphics[width=\textwidth]{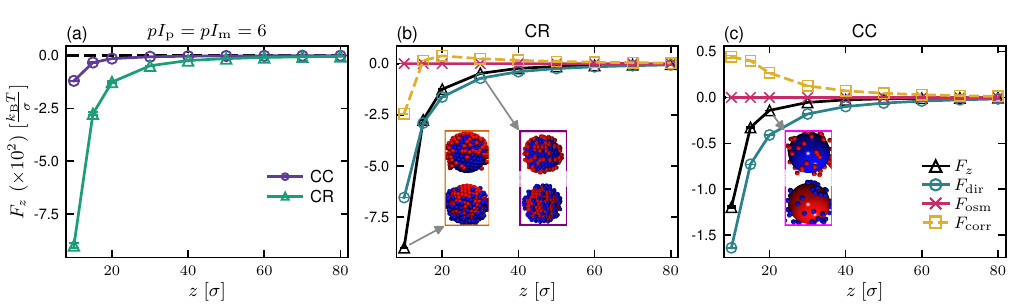}
    \caption{\textbf{Force decomposition for Asymmetric NPs}
    (a) Comparison of effective interaction force ${F}_{z}$ as a function of separation $z$ for CR and CC, and component wise breakdown of forces for (b) CR, insets are a snapshot of CR system at $z=10\sigma$, $z=30\sigma$, and (c) CC, inset is a snapshot of CC system at $z=20\sigma$, all at $pI_p=pI_m=6 (c\approx 10^{-6}$ in the electrolyte at acid and base dissociation constants $pK_a=pK_b=3$ for the surface sites.}
    \label{fig4}
\end{figure*}


We first present results from a system of asymmetric charged NPs configuration, in which one NP has acidic sites and the other has basic sites, as shown in Fig.~\ref{fig2}(a). Typical  simulation snapshot for CR and CC conditions corresponding to $pI_p=pI_m=3$ and a fixed separation ($z=15\sigma$) of $P_{1}$ and $P_{2}$ is shown in Fig.~\ref{fig2}(a). Notably, there seems to be a significant disparity between the two conditions, there are more free ions adsorbed on the NP surface in the CR condition compared to the CC condition, highlighting the pivotal role of electrostatic interactions and (de)ionization capacity of NPs. A quantitative description of the effect of CR is obtained by evaluating the fraction of adsorbed free ions ($f_{ad}$) relative to the total number of sites on the NP surface located within a $2\sigma$ range of the surface. In Fig.~\ref{fig2}(b), we plot the $f_{ad}$ as a function of solution concentration at a fixed NP separation of $15\sigma$, which further concretizes the capability of the CR model to capture close range interactions that may otherwise remain unrealized. Unlike CC, CR can redistribute its surface charge non-uniformly, causing ions in the solution to rearrange themselves near the NP surface. The spherical geometry of the NP further amplifies this uneven distribution of ions, thereby strengthening the mutual attraction force between $P_1$ and $P_2$\cite{ruixuan23mioncr, kandari26screencrzwit}. Figure~\ref{fig2} illustrates that  NPs strongly attracts the counterions, resulting in its enhanced adsorption over a wide regime of solution concentration in CR simulation. However, as the strength of the solution decreases, $f_{ad}$ is reduced in both CR and CC conditions. In fact, this occurs due to limited availability of mobile ions for absorption. These findings indicate that CR significantly enhances ion adsorption onto NPs analogous to the recent studies on PE adsorption onto a planar surface\cite{Yuan2024-ql} or NPs\cite{yadav26crpe}, and attraction between unlike charged PE pair\cite{Beyer2025}.

The interaction force between two asymmetric charged NPs as a function of $z$, is shown in Fig.~\ref{fig3} for high solution levels, and in Fig.~\ref{fig4} for the deionized condition.  
We first turn our attention to physiological solution condition. The distribution of adsorb mobile ions  over the surface of the $P_1$ and $P_2$  changes as the NPs approach one another, affecting the electrostatic surface potential. From small center-to-center separation to moderate separation, attraction in CR case is more prominent. This fact is quite visible from the snapshots presented in Fig.~\ref{fig2}(a) and adsorption behavior (Fig.~\ref{fig2}(b)). In high solution limit, we observe a $3$ to $4$-fold stronger attraction in CR than that observed in CC for closely placed NPs. As the separation between the NPs increases, the difference in total $F_z$ between the two conditions decreases rapidly, because the total forces in the CC and CR simulations converge at larger distances for both deionized water and physiological solution conditions, where $z >> R$. 
This behavior persist even for low ionic strength condition, see Fig.~\ref{fig4}. Our results indicate that at low ionic strength, the attraction between opposite charges at small $z$ is stronger than that observed for high concentration. This is associated with a reduced screening effect in dilute solution regime. In an ion-poor environment, amplified electrostatic attraction remains the dominant component of the interaction between the NPs, where we observed as much as $6$-fold stronger attraction within the same close separation range for CR compared to CC. Similar enhanced attraction is also observed between oppositely charged PE pair for a range of $pH$ values\cite{Beyer2025} and between asymmetrically charged NPs\cite{yuan22crzwit}. Furthermore, this interaction is stronger than that observed in ineine solutions\cite{yuan22crzwit}.

To illustrate the contributions of the various force components  we present Fig.~\ref{fig3}(b-c) and Fig.~\ref{fig4}(b-c), where the total force $F_z$ is divided into osmotic contribution ($F_{osm}$), direct force (Coulomb and non-electrostriction forces between  $P_1$ and $P_2$) and correlation force ($F_{corr}$). In Fig.~\ref{fig3}(b-c) and Fig.~\ref{fig4}(b-c) we plot different force components for CR and CC paradigms for all our considered values of $c \in [3,6]$. We can see that the effect of $F_{osm}$ is negligible and remains almost flat in all cases as the counterions prefer to adsorb onto the NPs as evidenced by the snapshot presented in Fig.~\ref{fig2}(a). Excluding the larger separation, we observe that at both low and high values of $c$, the $F_{cor}$ and $F_{dir}$ force components are the primary controlling factors. In ion rich environment, the repulsive nature of $F_{cor}$ persists in the CR and CC cases. However, the direct electrostatic force between $P_1$ and $P_2$ dominates here, and the attractive nature of $F_z$ arises from the contribution of the attractive direct force $F_{dir}$ and remains stronger for CR.

Now we focus on deionized solution case. We observe that except for $F_{corr}$, the nature of all other force components remains the same as what we observed in the case of the physiological solution solution (Fig.~\ref{fig3}(b-c)) . Furthermore, unlike CC picture, in CR the attractive nature of $F_z$ arises from the contribution of the attractive correlation $F_{corr}$ and $F_{dir}$ forces. The behavior of $F_z$ is somewhat more complex at small separations; the correlation force $F_{corr}$ which is repulsive for higher values of $c$ shifts from a repulsive to an attractive nature as the configuration changes from CC to CR at lower values of $c$. The attractive nature of $F_{corr}$ indicates that the NPs are fully ionized and situated in close proximity to one another, this leads to a significant redistribution of ion positions near the NP surfaces, thereby enhancing and facilitating the interactions between the NPs. As the distance between the NPs increases, the force components diminishes monotonically so as the total force $F_Z$.


\subsection{Symmetric Nanoparticles}

\begin{figure*}
    \centering
    \includegraphics[width=0.6\textwidth]{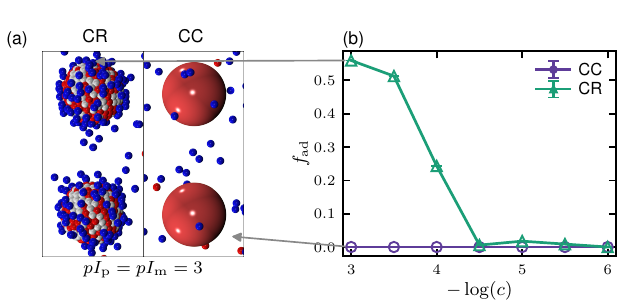}
    \caption{\textbf{CR compared with CC Baseline for Symmetric NPs}
    (a) Snapshots of the system at $pI_p=pI_m=3 (c\approx 10^{-3}$)  for a fixed  NPs separation of  $z=15\sigma$: CR (left) and CC (right)
    (b) Variation of adsorption fraction ($f_{ad}$) with electrolytic concentration $-\log(c)$ at separation $15\sigma$, at $pH=7.0$ with acid and base dissociation constants $pK_a=3$ and $pK_b=6$.}
    \label{fig5}
\end{figure*}

\begin{figure*}
    \centering
    \includegraphics[width=\textwidth]{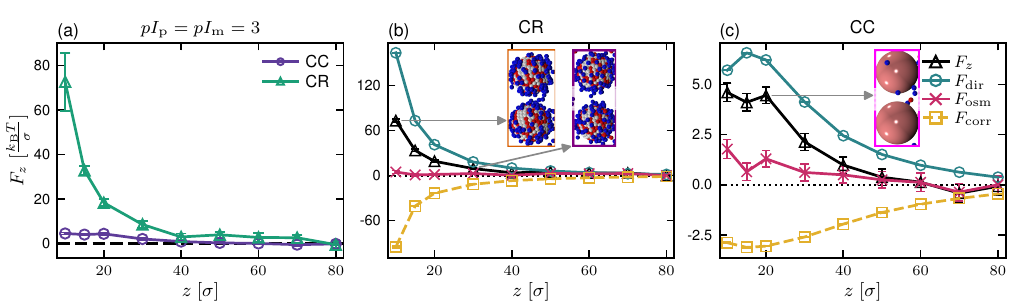}
    \caption{\textbf{Force decomposition for Symmetric NPs}
    (a) Comparison of effective interaction force ${F}_{z}$ as a function of separation $z$ for CR and CC, and component wise breakdown of forces for (b) CR, insets are a snapshot of CR system at $z=10\sigma$, $z=30\sigma$, and (c) CC, inset is a snapshot of CC system at $z=20\sigma$, all at $pI_p=pI_m=3$ in the electrolyte at acid and base dissociation constants $pK_a=3$, $pK_b=6$ for the surface sites.}
    \label{fig6}
\end{figure*}

\begin{figure*}
    \centering
    \includegraphics[width=\textwidth]{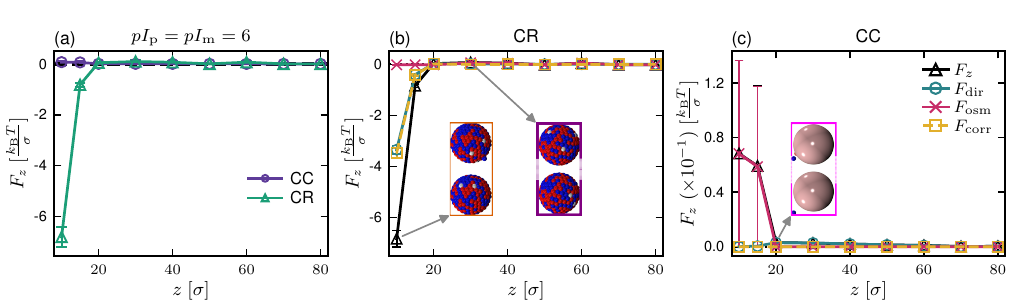}
    \caption{\textbf{Force decomposition for Symmetric NPs}
    (a) Comparison of effective interaction force ${F}_{z}$ as a function of separation $z$ for CR and CC, and component wise breakdown of forces for (b) CR, insets are a snapshot of CR system at $z=10\sigma$, $z=30\sigma$, and (c) CC, inset is a snapshot of CC system at $z=20\sigma$, all at $pI_p=pI_m=6 (c\approx 10^{-6})$ in the electrolyte at acid and base dissociation constants $pK_a=3$, $pK_b=6$ for the surface sites.}
    \label{fig7}
\end{figure*}

We now analyze the influence of CR on symmetrically charged NPs with $pK_a=3$ and $pK_b=6$, keeping all other parameters same as those mentioned earlier for the asymmetric case. Each NP surface is decorated with $128$ acidic sites and $128$ basic sites. Theses sites are randomly but uniformly distributed on the NP surafce. It is important to note here that NPs $P_1$ and $P_2$ are identical in all respect. Figure~\ref{fig5} displays the steady state simulation snapshot of $P_1$ and $P_2$ separated at $15\sigma$ for a fixed $c=3$. The Intriguing impact of CR compared to CC can be clearly observed in Fig.~\ref{fig5}(a), in contrast to CR result, CC simulation shows no absorption of mobile ions. The CR state promotes strong ion adsorption stemming from NPs surface dissociation equilibria. 

 We adopted similar method as we discussed for asymmetric case and calculated concentration dependence of $f_{ad}$ at fixed NPs distance of $15\sigma$ as shown in Fig~\ref{fig5}(b). When the solution contains more solution ions, the $f_{ad}$ value is high in CR simulation; it decreases rapidly with lowering concentration, and at very low concentrations, no adsorption is observed. Across the entire concentration range, $f_{ad}$ remains slightly lower compared to the case of oppositely charged NPs. Furthermore, as seen from the Fig.~\ref{fig5}(b) for any value of $c$, the adsorption remains approximately zero across the calculated range of center-to-center distance $z$ in CC case. To facilitate the discussion on the effect of CR on symmetrical NPs, we examine the pairwise interaction force between these two NPs. We plot the dependence of the total force $F_z$ between $P_1$ and $P_2$ on the distance for $c \in [3, 6]$ as shown in Fig.~\ref{fig6} and Fig.~\ref{fig7}, respectively. At high $c$, as illustrated in panel (a) of Fig.~\ref{fig6}, there is enhanced repulsion between charge regulated NPs. Due to saline solution, electrostatic screening reduces electrostatic interaction in both CR and CC simulations. Nevertheless, the repulsion is more due to the dynamical ionization capacity of the charge-regulated NPs compared to conventional CC at short separation consistent with the previous study on flat surfaces~\cite{Izzo2025}. At larger separation screening becomes effective and the difference between CC and CR diminishes. Figure~\ref{fig6}(b-c) indicates that the repulsive nature of $F_z$ involves a substantial  and qualitative contribution from the repulsion occurring between chemically identical and like-charged NP's direct force contribution $F_{dir}$. An attractive correlation force, $F_{corr}$, also contributes here due to the presence of charged entities on opposite sides of the midplane, as observed in the asymmetric case. However, the effect of $F_{dir}$ overshadows the contribution of $F_{corr}$, see Fig.~\ref{fig6}(b). Furthermore, as the NPs move further apart, the $F_{dir}$ and $F_{corr}$ components continuously decreases, directly affecting the total $F_z$ and causing $F_z$ to approach zero.

Finally we turn our attention to deionized solution. A number of interesting and subtle changes we can see in Fig.~\ref{fig7}. In the case of closely spaced NPs, where adsorption is negligible for both CC and CR, a transition from repulsive to attractive nature is observed in $F_z$ contrary to the high salinity CR scenario. As reflected in Fig.~\ref{fig7}(b), both Coulombic and correlation forces play a significant role at short distances. Unlike saline case, here both acidic and basic are dominated as evident in Fig~\ref{fig6}(b) and Fig.\ref{fig7}. When nanoparticles (NPs) are close to each other, significant opposite charges develop on their facing surfaces; however, as the distance between them increases, the interaction between the surface sites weakens. In this close $z$, solution induced attractive contribution are strong, leading to an attraction between pairs of like charged NPs. Several studies have reported similar trends of attraction characteristic between like charged particles, attributed to breaking of charge-inversion symmetry in the interparticle force\cite{ruixuan23mioncr, wang24qdrivenassembly, Wang2026}.  Furthermore, the electrosolvation attraction between negatively charged silica NPs can change its nature depending on the solution pH\cite{wang24qdrivenassembly}. We further note that as the NPs move away from each other, the attractive force between them weakens and subsequently, no interaction is observed.  On the other hand, across the range of investigated $c$, CC exhibits no interaction over a wide range of distances, which is naturally reflected in the adsorption profile as well, see Fig~\ref{fig7}(a). This is linked to the fact that CC NPs are either very weakly acidic or remain nearly neutral, as shown in the inset of Fig.~\ref{fig7}(c). To gain insight the $F_z$ behaviour in CC, we focus how the different  force components vary with the distance between $P_1$ and $P_2$ as displayed in Fig~\ref{fig7}(c). Except for very minor changes in the osmotic ($F_{osm}$) contribution, the contributions of all forces remain almost entirely constant and negligible with respect to $z$, thereby minimizing the impact of direct forces or other non-electrostatic effects. In the case of CC, this is linked to the previously shown $f_{ad}$ behaviour in Fig~\ref{fig5}(b) which illustrates the reluctance towards ion adsorption in a dilute solution.
 
\section{Conclusions}
\label{sec:conclusions}
Nanoparticle (NP) dressed with ionized functional groups are central for various biological processes and technical applications. Although classical DLVO theory has long served as a cornerstone of colloid science, yet it fundamentally relies on simplifying assumptions of constant charge (CC). In many real and complex situations, NPs dynamically adapt to their surrounding environment and regulate their stability and functional characteristics. By calculating the exact short-range forces between spherical charge regulated particles, this work directly addresses these classical limitations.

In this work, we quantified the pairwise interaction and ion adsorption behavior between spherical CR NPs using explicit-ion molecular dynamics (MD) coupled with Monte Carlo (MC). By evaluating both chemically complementary and chemically similar surface configurations against a CC baseline, we systematically isolated the role of surface CR in modulating short-range electrostatics. We also compared our findings for the CR model with a CC baseline established based on the average surface charges and $c$ during equilibrium state of CR progressions.

We first simulated a pair of CR spherical NPs submerged in a monovalent electrolyte. The interactions were constituted by the Columbic potential and the non-electrostatic potential. Two specific configurations were studied. In the asymmetric configuration, the surfaces of the pair of NPs were chemically complementary to each other; one decorated entirely with acidic surface sites, and the other, basic. For the symmetric configuration, the surfaces were embedded with both the species of surface sites, acidic as well as basic, in equal proportion, and uniform random distribution, while the surfaces on the whole were identical to each other.

{We observed that in the asymmetric configuration of NPs, the number of adsorbed ions on the surface decreased steadily with solution density, while a similar yet diminished trend was observed in the CC case as well. Sensitive to environmental cues, CR amplifies the short range attractive interaction between NPs up-to four fold for ion rich and up to six fold at low concentration, compared to CC. From the component wise force analysis, we inferred that the direct interaction between the fluctuating charge densities on the CR surfaces was a major influence on the enhanced interaction. The enhancement was also favored by the correlational component, and even more so at high solution density, due to the component being directly supported by the higher adsorption of the free ions. The osmotic repulsions, although not entirely non-existent, were diminished significantly in comparison to the direct NPs  interactions.

Within the same framework, chemically identical NPs showed a quite distinct behaviour. The number of adsorbed ions sharply fell with ionic strength in CR, while remaining close to zero in CC. The net charge in the CC baseline also remained small. As a result, at a high solution strength, we observed enhanced repulsions compared to CC, up to about 10-12 times stronger at very small inter-surface separation of NPs. Remarkably, in deionized solution, the interaction transitioned to a weakly attractive nature in CR for electrically like NPs, while in CC this transition did not occur. Component-wise analysis of the interaction indicated again the major influence of the direct component on the NP interaction. However, due to the surfaces hosting an equal number of complementary charged species, the direct interaction could not obscure the influence of forces originated between entities located on opposite sides of midplane as much as in the case of unlike charged NPs. In contrast to the asymmetric configuration, the influence of each component had a more complex effect on the net interaction. The correlational components remained fairly attractive, while CR did not enhance the osmotic repulsion. The direct component itself transitioned from strong repulsions to weak attraction at close separations with a lowered $c$. Correlational component reduced the CR enhancement at high concentration, while favoring it at low ionic strength.

Furthermore, when the separation between NPs ranges from moderate to large, the interaction forces in both CR and CC paradigms decreases, and ultimately, the difference between CR and CC vanishes within the calculated concentration range. A similar trend is observed in the case of symmetric charges. We expect that our results obtained
in dilute solution may pave the way to more complex phenomena, including like charge self assembly of nanosized particles to macromolecules.

In conclusion, CR is not merely a quantitative correction to CC. Not only did CR result in significant close range enhancement of interaction, it also qualitatively influence the nature of the interaction between NPs. Our study uses a simplified setup, with rigid and spherical NPs in a monovalent solution to isolate the effects of charge regulation. The addition of multivalent solution and surface shape modulations will provide a better representation of a large number of real experimental situations. Incorporating system-specific details into future studies will be essential in providing deeper insights into these dynamic interactions.

\section{Acknowledgment}
We would like to pay tribute to the late Professor Podgornik. He was a pioneer in the fields of electrostatic interactions, biophysics, and soft matter. His inspiring
guidance profoundly shaped our understanding of charge-regulation processes and laid the foundation for the work presented here. 

ST acknowledges the fellowship provided by the Ministry of Education (MoE), Government of India. SK acknowledges the financial support received from the Anusandhan National Research Foundation, India ( EEQ/2023/000676) and IIT Jodhpur for a research initiation grant (I/RIG/SNT/20240068).


\FloatBarrier

\bibliography{Manuscript} 

\end{document}